\documentclass[twocolumn]{aastex63}

\usepackage{tipa}
\usepackage{balance}

\submitjournal{ApJ Letters}

\shorttitle{Quenching, Mergers and Age Profiles}
\shortauthors{Pathak et al.}
\graphicspath{{./}{figures/}}

\begin{document}

\title{Quenching, Mergers and Age Profiles for $z=2$ Galaxies in IllustrisTNG}

\correspondingauthor{Debosmita Pathak}
\email{pathakde@grinnell.edu}

\author[0000-0003-2721-487X]{Debosmita Pathak}
\affiliation{Grinnell College \\
1115 8th Avenue \\
Grinnell, IA 50112, USA}

\author[0000-0002-5615-6018]{Sirio Belli}
\affiliation{Center for Astrophysics \textpipe Harvard \& Smithsonian \\
60 Garden Street \\
Cambridge, MA 02138, USA}

\author[0000-0001-6260-9709]{Rainer Weinberger}
\affiliation{Center for Astrophysics \textpipe Harvard \& Smithsonian \\
60 Garden Street \\
Cambridge, MA 02138, USA}





\begin{abstract}
Using the IllustrisTNG cosmological galaxy formation simulations, we analyze the physical properties of young quiescent galaxies at $z=2$ with stellar masses above $10^{10.5} M_\odot$. This key population provides an unaltered probe into the evolution of galaxies from star-forming to quiescent, and has been recently targeted by several observational studies. Young quiescent galaxies in the simulations do not appear unusually compact, in tension with observations, but they show unique age profiles that are qualitatively consistent with the observed color gradients. In particular, more than half of the simulated young quiescent galaxies show positive age gradients due to recent intense central starbursts, which are triggered by significant mergers. Yet, there is a sizable population of recently quenched galaxies without significant mergers and with flat age profiles. Our results suggest that mergers play a fundamental role in structural transformation, but are not the only available pathway to quench a $z=2$ galaxy.
\end{abstract}



\section{Introduction} \label{sec:intro}
One of the main challenges for models of galaxy evolution is to explain the properties of massive quiescent galaxies and to understand the mechanisms responsible for quenching their star formation.
The observed diversity in the structure and stellar populations of quiescent galaxies suggests that there are multiple quenching pathways \citep[e.g.,][]{10.1093/mnras/stu327}. One way to explore the physical processes related to quenching is to focus on the population of galaxies that recently became quiescent. Over the last decade, observational studies discovered a relatively large number of these young quiescent systems, sometimes called \emph{post-starburst galaxies}, at high redshift \citep[e.g.,][]{wild09, 2012ApJ...745..179W}. 
While old quiescent galaxies at high redshift are remarkably small, post-starburst systems \textit{appear} even more compact  \citep{2012ApJ...745..179W, 2015ApJ...799..206B, 2016ApJ...817L..21Y, 2017MNRAS.472.1401A, 2018ApJ...868...37W}. However, recent studies suggest that the difference in half-light radii between young and old quiescent galaxies at $z\sim2$ is actually due to a difference in color gradient rather than in physical size \citep{suess20}. This may be due to a central population of young, blue stars, but direct measurements of the age profiles in post-starburst galaxies can currently be obtained only at $z<1$ \citep{2020MNRAS.497..389D, setton20}.

These results raise interesting questions for current models of galaxy quenching. Some theoretical scenarios propose that massive quiescent galaxies form via a compact starburst triggered by violent disk instabilities (including mergers) that drive cold gas into the center \citep{2014MNRAS.438.1870D}. A scenario of this type can qualitatively explain the observed sizes and colors of post-starburst galaxies at $z\sim0.8$ \citep{wu20}. Hydrodynamical simulations have enabled detailed studies of the relation between gas-rich processes, central starbursts, and quenching \citep{zolotov15, tacchella16}. However, these zoom-in simulations do not include feedback from active galactic nuclei (AGN), which is widely thought to be a crucial component of galaxy quenching \citep{2006MNRAS.365...11C, 2006MNRAS.370..645B}, and it is therefore unclear whether they are able to reproduce realistic properties for the global population of galaxies.
Other studies include AGN feedback and focus on understanding the interaction between galaxy mergers, quenching and AGN feedback \citep{pontzen17, sanchez21}, AGN activity \citep{sharma21} or structural evolution \citep{choi18}.
On the other hand, large-scale cosmological simulations are ideal for capturing the vast diversity in galaxy populations and derive statistically significant results relevant to observational surveys. Only recently, the physical fidelity and resolution of these simulations became sufficiently good to study the spatially resolved structure of galaxies. Yet, striking a balance between simulated volume, i.e. sample size, and resolution remains an important aspect. In this Letter\footnote{
The code used in our analysis is available on GitHub under a GPL-2.0 license \url{https://github.com/pathakde/Pathak_2021}; a copy of these data were deposited in Zenodo \citep{zenodo}}, we use the TNG-100 simulation from the IllustrisTNG suite, which provides the optimal balance between resolution and sample size, to study stellar age gradients as signature of quenching. We focus our analysis on the galaxy population at $z=2$, a key epoch marking both the peak of the cosmic star formation rate \citep{2014ARA&A..52..415M}, and the time when quiescent systems begin to dominate the high end of the mass function \citep{muzzin13}.

\section{Simulation data and sample selection} \label{sec:simulation}

\subsection{IllustrisTNG} \label{sec:TNG}

The IllustrisTNG project \citep{2018MNRAS.480.5113M, 2018MNRAS.477.1206N, 2018MNRAS.475..624N, 2018MNRAS.475..648P, 2018MNRAS.475..676S} is a series of large-scale cosmological magnetohydrodynamical simulations that model galaxy formation and evolution, including cooling, star formation, stellar feedback and AGN feedback. We use data from TNG100-1, the highest resolution realization of the TNG100 series using the web-based API \citep{nelson19}. The side length of the simulation volume is $75 h^{-1} \approx 100 \text{ Mpc}$ (comoving), the baryon mass resolution is $1.4 \times 10^6 M_\odot$ per particle, the gravitational softening length for dark matter and star particles at $z=2$ is $0.5$~kpc (proper), and the adaptive softening of the gas component has a minimum of $62$~pc (proper). The \citet{Ade2016} cosmological parameters are adopted. Gravitationally bound structures are identified using the SUBFIND algorithm  \citep{2001MNRAS.328..726S} and are linked across snapshots to generate merger trees with the SUBLINK algorithm \citep{2015MNRAS.449...49R}. All TNG model parameters used are described in \citet{2018MNRAS.475..648P}. Details on the physical models used in TNG, in particular the models for star formation, stellar and AGN feedback, can be found in the two TNG method papers \citep{2017MNRAS.465.3291W, 2018MNRAS.473.4077P}.

The simulated galaxy population in TNG100 has been extensively analyzed and compared to observations in the literature. Most relevant to this work, \citet{genel18} found the sizes of simulated galaxies to be in rough agreement with observations (also at $z=2$, the focus of the present study). Additionally, \citet{weinberger18} traced back quiescent galaxies from $z=0$
to the time of quenching and found that during quenching, most galaxies exhibit increased levels of kinetic mode AGN feedback, and only a small sub-population show substantial energy injection via luminous AGN. This implies that in IllustrisTNG, most galaxies do not quench via luminous AGN, and hints towards kinetic AGN feedback being responsible for quenching.

\subsection{Sample Selection} \label{sec:sample}

\begin{figure*}[htbp]
    \centering
    \includegraphics[width=1.0\textwidth]{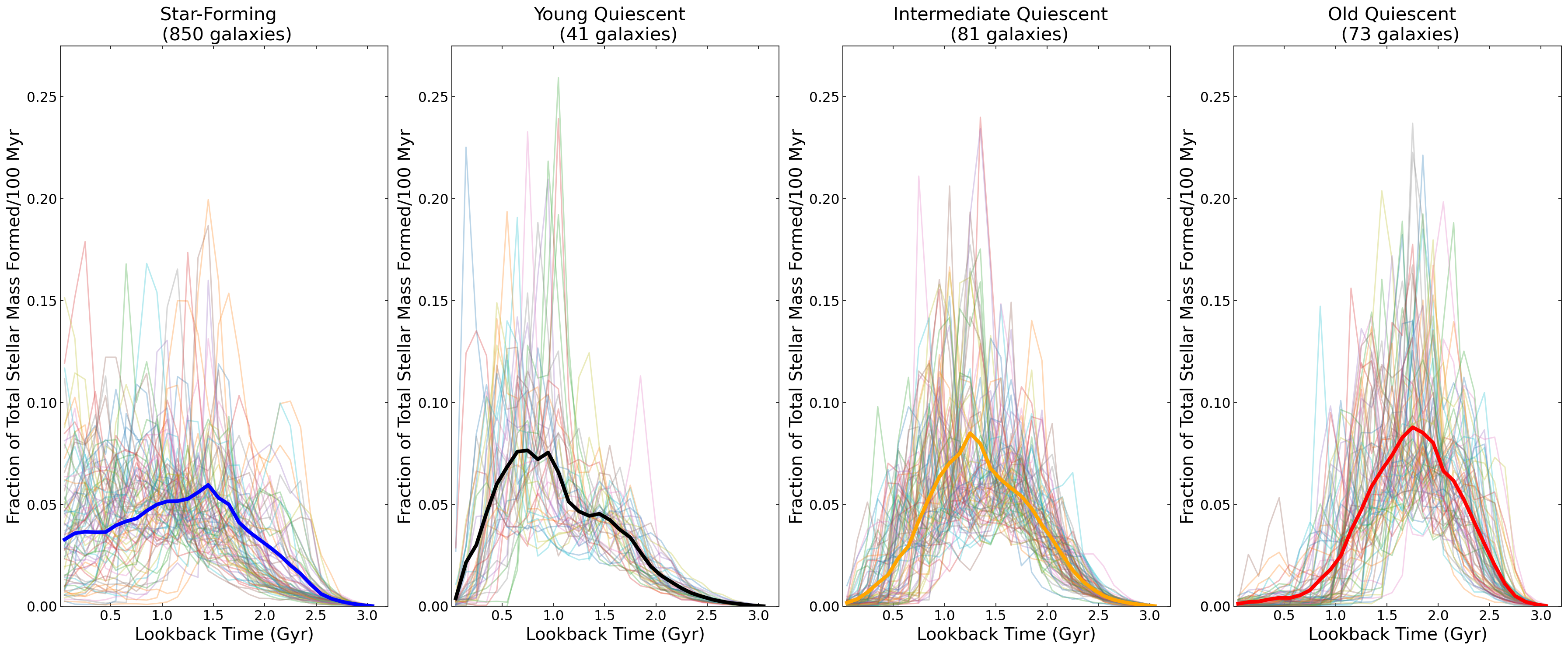}
    \caption{Star formation histories for the four galaxy populations. For the star-forming population we only show $80$ randomly drawn galaxies, for clarity. The curves are binned over $100$ Myr and normalized by the total mass formed \textbf{by $z=2$}. In each panel, a thick line marks the median trend. The data and software to generate this Figure are available on Zenodo: \url{10.5281/zenodo.5093727} (GitHub: \url{https://github.com/pathakde/Pathak_2021)}.} 
    \label{fig:SFH}
\end{figure*}

We focus our analysis on the $z=2$ snapshot of the TNG100 simulation, for which we select all 1045 galaxies with stellar mass $M_* \geq 10^{10.5} M_{\odot}$. We split this sample into 850 star-forming galaxies and 195 quiescent galaxies using a threshold in specific star formation rate (averaged over the last 100 Myr) of $10^{-10.5}$ yr$^{-1}$, which is an order of magnitude below the main sequence at $z\sim2$ \citep{speagle14}. We further split the quiescent population into subsamples according to the mass-weighted average stellar ages. The main focus of this work is on the 41 \emph{young quiescent} galaxies with an average age younger than 1.15 Gyr. We choose this age threshold because it yields a number density of $1.5\times10^{-5}$ Mpc$^{-3}$, which is roughly equal to the observed number density of massive post-starburst galaxies at high redshift \citep{2016MNRAS.463..832W, 2019ApJ...874...17B}. Selecting the sample via such abundance matching reduces biases due to possible differences in quenching times between simulations and observations\footnote{We also verified that our conclusions do not change with a threshold of $1$~Gyr, as used in \citet{suess20}.}.
We also define the two control populations of intermediate-aged quiescent galaxies (mass-weighted average stellar age between $1.15$ and $1.5$ Gyr, 81 systems), and old quiescent galaxies (mass-weighted average stellar age older than 1.5 Gyr, 73 systems). 

The star formation histories for each of the four populations are illustrated in Figure \ref{fig:SFH}. Most young quiescent galaxies show a peak in star formation between 0.5 and 1 Gyr of lookback time, followed by rapid quenching. The star formation histories of intermediate and old quiescent galaxies feature pronounced peaks at increasingly early cosmic times.

\section{Structural properties of galaxies} \label{sec:results}

\subsection{Sizes} \label{sec:sizes}

\begin{figure*}[htbp]
    \centering
    \includegraphics[width=1\textwidth]{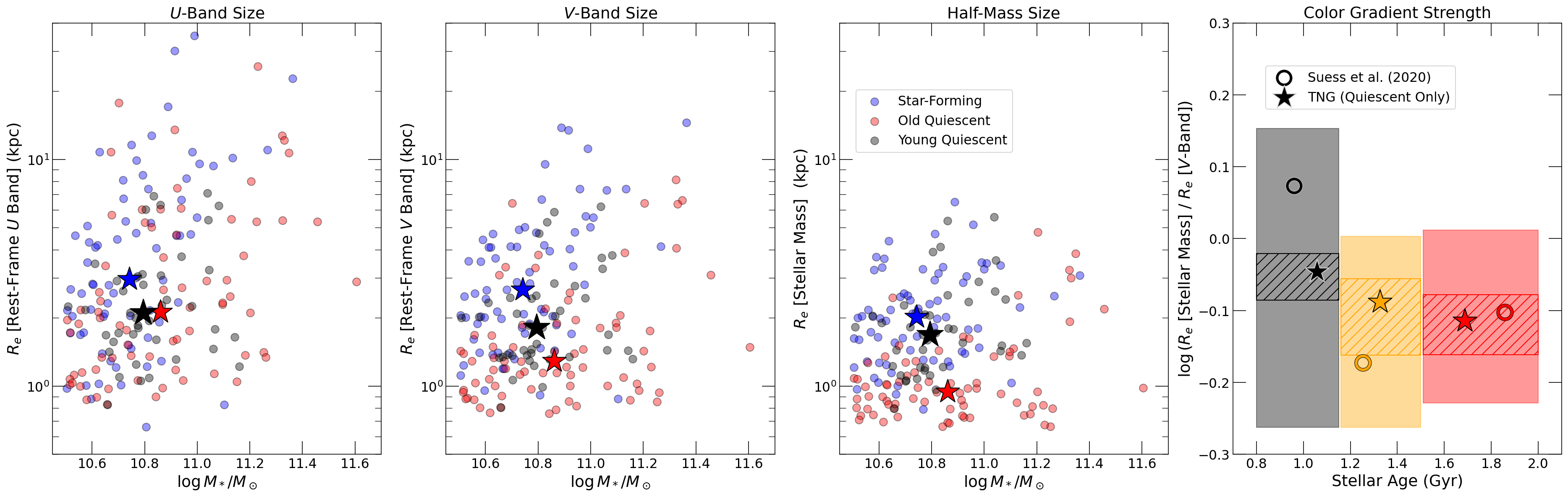}
    \caption{The first three panels show the effective sizes calculated in the rest-frame $U$ band, rest-frame $V$ band, and using the stellar mass distribution, as a function of total stellar mass. Galaxies are color-coded according to the populations defined in Section~\ref{sec:sample}. For the star-forming population we only show $80$ randomly drawn galaxies; we also omit the intermediate quiescent population from the first three panels for clarity. Stars mark the median values for each population. The last panel shows the ratio of half-mass and half-light (in the rest-frame $V$ band) size as a function of stellar age, for the three subsamples of quiescent galaxies. Star symbols mark the medians for the TNG galaxies, and empty circles mark the medians for the \citet{suess20} observed sample at $1.5 < z < 2.5$. The 30th-70th percentile regions for both the simulation (hatched) and observations (shaded) are shown. The data and software to generate this Figure are available on Zenodo: \url{10.5281/zenodo.5093727} (GitHub: \url{https://github.com/pathakde/Pathak_2021)}.}
    \label{fig:halfmass}
\end{figure*}

In order to explore the relation between quenching and structural transformation, we calculate the effective radii of the simulated galaxies. We retrieve the multi-band photometry generated with the \citet{bruzual03} library for each stellar particle, and calculate the three-dimensional half-light size of each galaxy in the rest-frame $U$ and $V$ bands. For the purpose of this work we ignore dust attenuation, and assume that the luminosity of a stellar particle depends only on stellar age, metallicity and mass \citep{vogelsberger13}. The first two panels of Figure~\ref{fig:halfmass} show the $U$-band and $V$-band effective sizes as a function of stellar mass for our sample, color-coded by the galaxy populations as defined in the previous section. Generally speaking, we find that in TNG100 star-forming systems are the largest at fixed stellar mass. This aspect was studied in \citet{genel18} and found to be in agreement with observations. Within the quiescent population, we find that young galaxies are larger than old galaxies in the $V$ band, and about the same size in the $U$ band. This is in tension with observations at $0.5 < z < 2.5$, in which young quiescent systems are found to be the most compact type of galaxy \citep{2012ApJ...745..179W, 2015ApJ...799..206B, 2016ApJ...817L..21Y, 2017MNRAS.472.1401A, 2018ApJ...868...37W}. 

\begin{figure*}[htbp]
\includegraphics[width=1.0\textwidth]{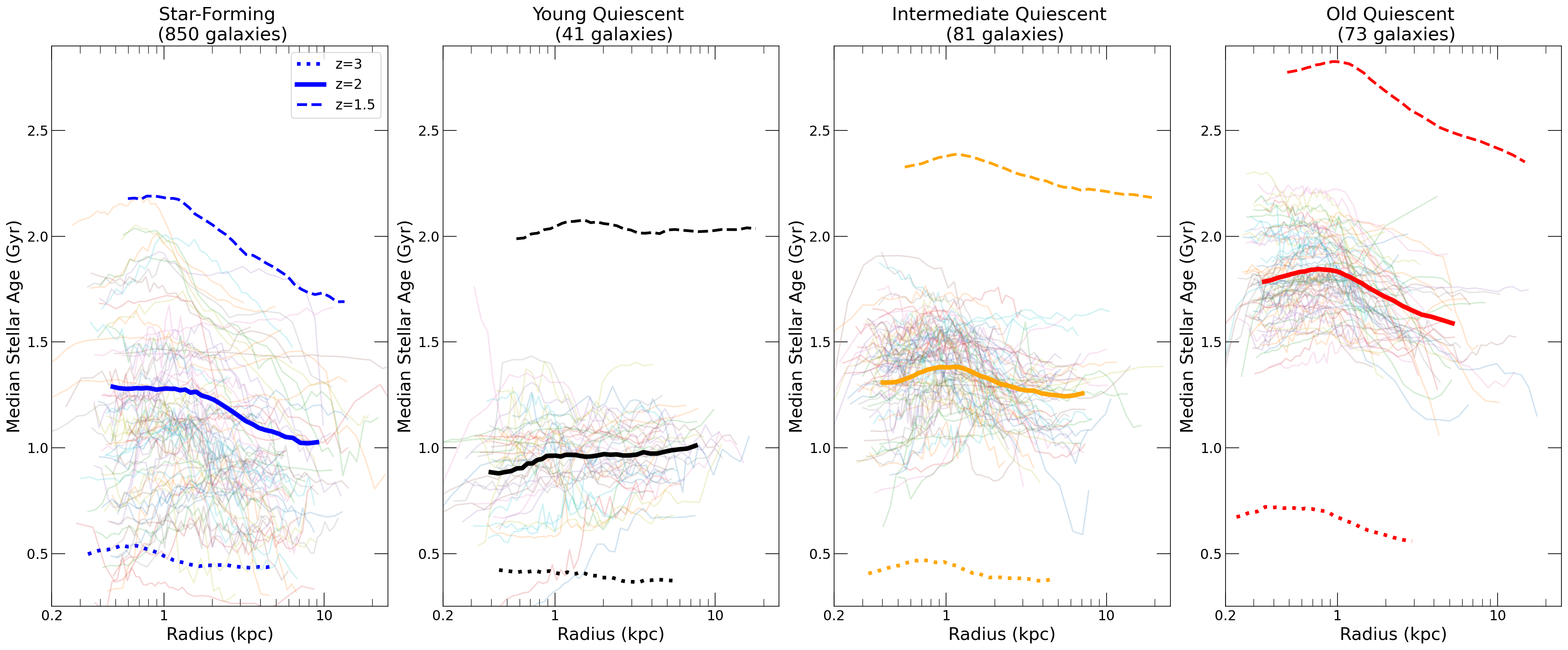}
\caption{Radial age profiles for the four galaxy populations (thin lines). For the star-forming population we only show 80 randomly drawn galaxies, for clarity. The median for each population is shown in thick unbroken lines. The median for the main progenitors at $z=3$ (thick dotted line) and the descendants at $z=1.5$ (thick dashed line) are also shown. The data and software to generate this Figure are available on Zenodo: \url{10.5281/zenodo.5093727} (GitHub: \url{https://github.com/pathakde/Pathak_2021)}.}
\label{fig:age_profiles}
\end{figure*}

Since the size difference between young and old quiescent galaxies depends on the band in which the effective radii are measured, these two galaxy populations must have substantially different color gradients. More precisely, the young quiescent population must have a nearly flat color gradient since its sizes are roughly the same in both bands, whereas older galaxies are more compact in the redder filter and must therefore have strong negative gradients (i.e., redder in the center). These results are in qualitative agreement with recent observational findings based on multi-band space-based imaging \citep{suess20}.
We also calculate the half-mass sizes, which can be considered as the ``true'' size measurements, and we still find a tension between simulations and observations. As shown in the third panel of Figure~\ref{fig:halfmass}, in TNG100 young quiescent galaxies are much larger than old systems, while the half-mass size of observed galaxies seem to be roughly independent of their age \citep{suess20}.
Finally, the last panel of Figure~\ref{fig:halfmass} shows the ratio between the half-mass and half-light size, which is a measure of the color gradient, as a function of stellar age. Our sample (star symbols) follows a weak but clear trend of decreasing color gradient with stellar age, which is in qualitative agreement with the observational measurements of \citet[][circles]{suess20}. The observed gradients feature a substantially larger scatter than the TNG measurements, but this is explained at least in part by the effect of observational uncertainties and the small number of quiescent galaxies available in \citet[][]{suess20}. 

We conclude that in our simulated sample, young quiescent galaxies are larger than in  observations, but the color gradient trend is correctly reproduced, suggesting that TNG100 is able to capture, at least partially, the physical processes responsible for the formation of quiescent galaxies at $z=2$. Color gradients are an important observational property of galaxies, and reflect gradients in stellar age, metallicity, and/or dust reddening. The TNG100 simulation does not include dust, and we checked that the stellar metallicity profile is approximately the same for the four galaxy populations. This means that the color gradients in our simulated sample are mainly driven by gradients in the stellar ages.

\subsection{Age profiles} \label{sec:age_profiles}

\begin{figure*}[htbp]
\centering
\includegraphics[width=0.6\textwidth]{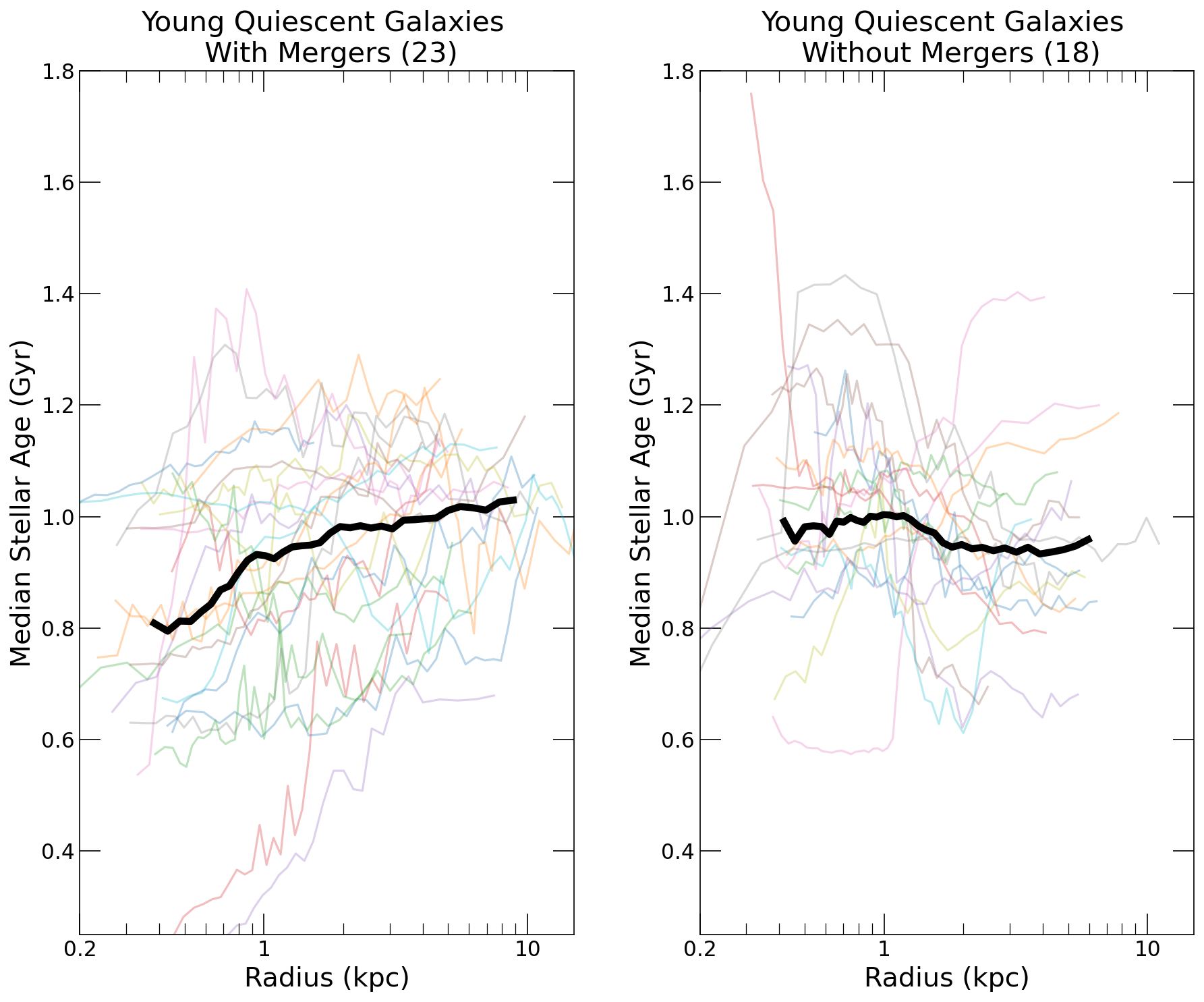}
\caption{The age profiles of young quiescent galaxies, with separate panels for galaxies with at least one significant merger that contributes at least $10\%$ of the $z=2$ stellar mass (`with merger': left panel) and those without such a merger (right panel). The data and software to generate this Figure are available on Zenodo: \url{10.5281/zenodo.5093727} (GitHub: \url{https://github.com/pathakde/Pathak_2021)}.}
\label{fig:PSB_merger_ageprofiles}
\end{figure*}

We construct the age profile for each galaxy by grouping the star particles in radial bins, each containing 2\% of the total amount of stars, and calculating the median stellar age in each bin. The results are shown in Figure \ref{fig:age_profiles}, separately for each of the four galaxy populations. Thick lines mark the population median trends, calculated using the star particles of all galaxies collectively (this increases the weight of individual massive galaxies and compensates for their lower number).
The age profiles of the three quiescent populations show a systematic vertical shift, which is simply due to the selection in median ages used to define our subsamples. Partly because of this, each quiescent population appears uniform, whereas the star-forming population, which does not have an age cut, shows a greater diversity. More importantly, there is a clear trend in the \emph{slope} of the age profiles: most galaxies have a negative slope between $0.8$ and $8$~kpc with the notable exception of the young quiescent population, which features a slightly positive slope.
Figure~\ref{fig:age_profiles} thus confirms that the age profiles are the cause for the difference in the mass-size relations measured in the rest-frame $U$ versus $V$ band. Old systems have an older center, which is therefore redder, while for young quiescent galaxies the center is younger and bluer than the outskirts.

To investigate the evolution of the age profiles, we follow each galaxy back in time to $z=3$ and forward in time to $z=1.5$, and we plot the median age profile for each of the four populations in Figure~\ref{fig:age_profiles}. The slope of the age profile tends to become increasingly negative with time across all populations. Moreover, the positive slope seen for young quiescent galaxies is acquired and lost in a short period of time, and the $z=3$ progenitors of these galaxies have age profiles that are effectively indistinguishable from those of star-forming galaxies. This indicates that the unique age profiles found in young quiescent galaxies at $z=2$ are not inherited from their progenitors: some event, possibly connected with the quenching process, is required.

\section{The role of mergers} \label{sec:mergers}

To test whether the unusual age profiles of young quiescent galaxies are associated with galaxy mergers, we further split our samples of galaxies by the presence or absence of a significant merger. Using the merger identification algorithm of \citet{2015MNRAS.449...49R}, we define `with merger' those galaxies whose most massive merger contributed at least $10\%$ of their $z=2$ stellar mass. The rest of the population are labeled `without merger.'
We use this merger criterion to split the population of young quiescent galaxies into two groups, and we show their age profile separately in Figure~\ref{fig:PSB_merger_ageprofiles}. Young quiescent galaxies which experienced at least one merger show strongly positive age gradients, while young quiescent galaxies that evolve without such a significant merger have flat age profiles. The median time since the most significant merger for young quiescent galaxies is 0.61 Gyr. This corresponds to the peak of star-formation and the onset of quenching in most young quiescent galaxies, as shown in Figure \ref{fig:SFH}.

\begin{figure*}[htbp]
\centering
\includegraphics[width=1.0\textwidth]{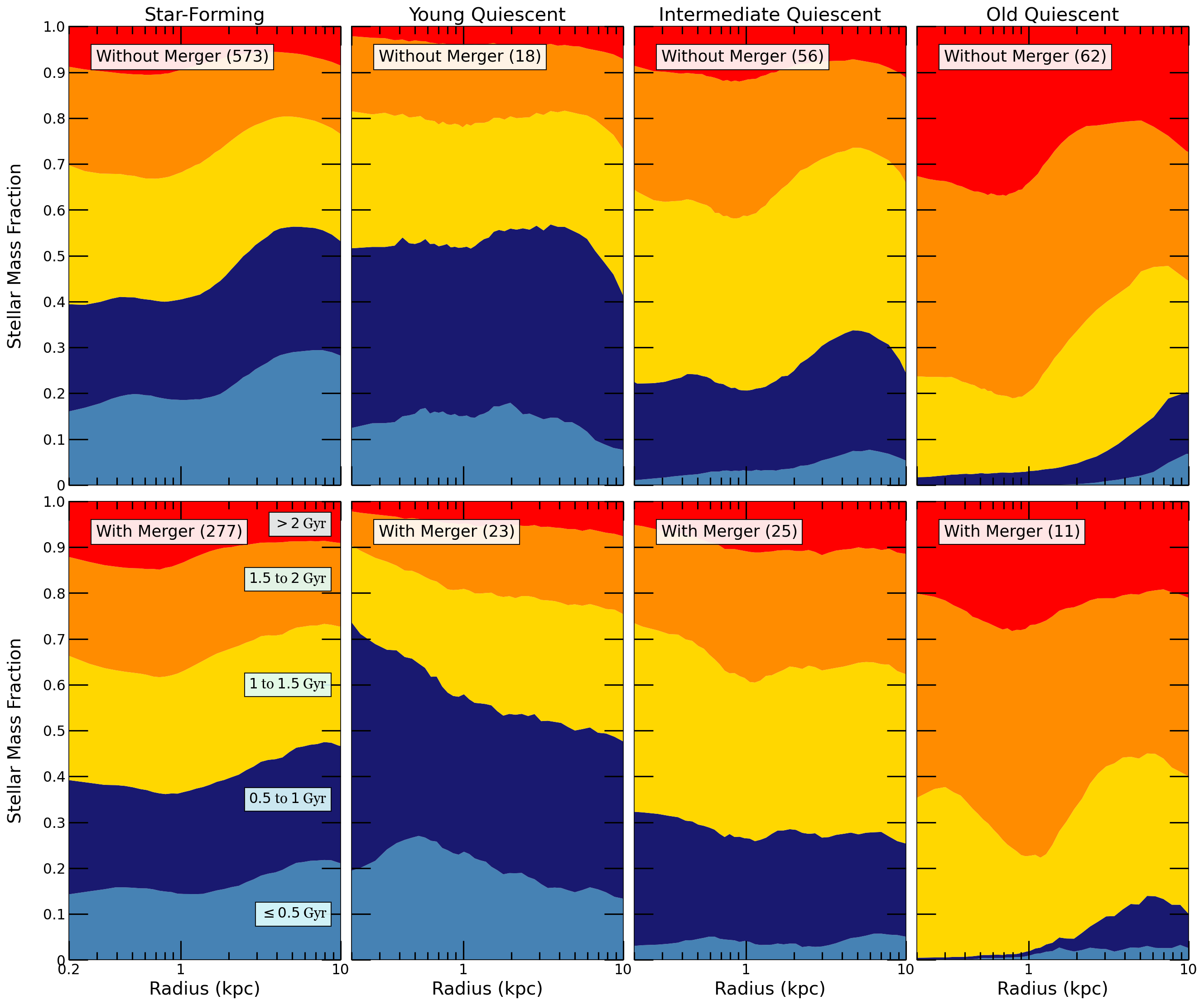}
\caption{Stacked normalized histograms showing the radial distribution of stellar ages in each galaxy population. All stellar particles of all galaxies in a given population are split into five age bins; the colored bands show the fractional contribution of each bin to the total stellar mass, as a function of radius. Galaxies are divided according to whether they experienced (bottom panels) or did not experience (top panels) a significant merger prior to $z=2$. The number of galaxies in each category are indicated in parentheses. The data and software to generate this Figure are available on Zenodo: \url{10.5281/zenodo.5093727} (GitHub: \url{https://github.com/pathakde/Pathak_2021)}. }
\label{fig:age_norm_SFH}
\end{figure*}

Thus, it appears that positive age gradients are caused by galaxy mergers. To further explore this process, in Figure \ref{fig:age_norm_SFH} we show the normalized stacked histograms of stellar ages as a function of radius in each galaxy population.
Radial percentile bins consistent with those in previous figures are used for binning stellar particles. At each radial bin, we show the overall stacked distribution of stellar ages, normalized by the total stellar mass within the bin to account for the high concentration of stars close to the center.
In most galaxy populations we find that the relative contribution of stars formed in the last Gyr peaks at a radius between 3 and 10 kpc, leaving mostly older stars in the central 3 kpc. The only exception is represented by the population of young quiescent galaxies that experienced a merger: in these systems, young stars peak at the galaxy center, and dominate the inner 3 kpc. 

This represents conclusive evidence for an intense and centrally concentrated starburst that is associated with a merger event.
The positive age gradients in the population of young quiescent galaxies are therefore caused by the formation of young stars in the center following a merger. 
Importantly, this process does not take place in every young quiescent galaxy, meaning that mergers are not required in order to trigger the quenching of massive galaxies. However, since the majority of young quiescent galaxies do experience mergers (23 out of 41, $\sim 60\%$), this process has a measurable effect on the median age profile of the population.

\section{Discussion} \label{sec:conclusion}

\subsection{Comparison to Observations}

We find that the age gradients are slightly positive in young quiescent galaxies and strongly negative in older systems. Assuming that dust attenuation and metallicity do not vary strongly with radius in quiescent galaxies, our results can be directly translated into color gradients. We thus confirm the observational results of \citet{suess20}, who showed that young quiescent galaxies are bluer (i.e., younger) in the center compared to older systems. Ideally, we would compare our analysis to direct measurements of age gradients, so that uncertainties due to the treatment of dust and metallicity (both in the simulations and in the observations) can be avoided. However, such measurements are extremely challenging, as they require deep and spatially resolved spectroscopy. Recent studies were able to measure age gradients in post-starburst galaxies at intermediate redshift ($z\sim0.7$), but a consensus has not emerged yet: while \citet{2020MNRAS.497..389D} report sightly positive age gradients, \citet{setton20} find flat age profiles. At higher redshift it is currently impossible to resolve quiescent galaxies, but direct measurements can be obtained in rare cases by taking advantage of gravitational lensing. So far only two such galaxies at $z\sim2$ have been studied in detail, and both feature flat age gradients \citep{jafariyazani20, akhshik20}. However, the stellar ages of these systems are relatively old, and we would not classify them as young quiescent galaxies.

\subsection{The Relationship between Mergers and Quenching}

Our results show that mergers trigger central starbursts and fundamentally change the age profile for the majority of young quiescent galaxies. Yet, we also find a sizable population of recently quenched galaxies that are quenched without mergers nor starbursts, and that have flat age profiles. This result suggests the existence of different pathways to quenching for $z=2$ massive galaxies, at least in the TNG simulations. By using the key signature of positive age gradients, observations can in principle identify those galaxies in which quenching was associated with a merger. By measuring how frequent positive age gradients are, and what is the observed diversity in age profiles among recently quenched galaxies, it will be possible to accurately establish the role played by mergers in the real universe. 
Using the more indirect probe offered by color gradients, \citet{suess21} conclude that high-redshift galaxies must follow a range of pathways to quenching, some of which may include a central starburst.
Direct measurement of age gradients in large samples at $z\sim2$ will hopefully be obtained by upcoming facilities such as the James Webb Space Telescope.

It is important to note that in this study we are exclusively referring to the immediate effects of mergers in cosmic noon galaxies \citep[similar to][]{pontzen17}, which are different from the effects of mergers over cosmic time \citep[e.g.][]{choi18}. Indeed the latter effect on the distribution of stars will cause the discussed signatures to fade away, highlighting the need to study recently quenched galaxies to learn about the quenching process.

\subsection{Inside-out or Outside-in Quenching?}

One additional result visible in Figure 5 is the fact that in both star-forming and older quiescent galaxies the population of young stars does not peak at the center, but around 5 kpc. This is likely due to different physical reasons (in-situ formation for star-forming systems versus accretion for quiescent systems), but has the result that all these galaxies must experience an \emph{inside-out} growth, and are becoming larger with time. This is in general agreement with observations, however we caution that our age profile measurements should not be directly compared to the profile of specific star formation rate, which is what is usually measured at high redshift \citep{tacchella15, nelson16}. In fact, one could say that the population of galaxies undergoing a merger-induced central starburst is experiencing ``outside-in quenching'', even though the specific quenching mechanism in TNG is, by construction, placed at the center of the galaxy. 
Moreover, the central starburst is likely responsible for the formation of a bulge  \citep{tacchella19}, which may be consistent with observations of dust-obscured star-formation at $z\sim2$ \citep{tadaki20}. After this bulge is formed, its additional mass concentration will naturally lower the specific star formation rate in the center \citep{abramson14}, which is in fact observed in TNG \citep{nelson21}. Such a system may then be assigned an ``inside-out quenching'' label when in fact it just completed an ``outside-in'' process. The confusion is partly due to the fact that the specific quenching mechanism (in this case, AGN feedback) is only one among many physical processes that are responsible for the galaxy transformation.

\subsection{Limitations and future improvements}

While we consider the merger--age gradient correlation a robust prediction of the model, there are remaining discrepancies: unlike in observations, in IllustrisTNG young quiescent galaxies are larger than old quiescent galaxies. There are a variety of possible reasons for this:
    \begin{itemize}
        \item \emph{Dust:} The emitted radiation might be partially absorbed by dust, leading to different observed sizes. Since we are comparing galaxies at very different stages of their evolution, with vastly different star formation histories, quasar activities and gas properties, it is likely that the resulting dust properties in these galaxies are very different and need to be taken into account to obtain unbiased size measurements.
        \item \emph{Resolution:} The size of the young stellar populations is of the same order as the gravitational softening length ($0.5$~kpc), thus even if stars would initially form more centrally concentrated, the limited mass resolution would lead to substantial numerical inaccuracies and prevent such a centrally peaked density distribution to be accurately modeled \citep{ludlow2019}.
        \item \emph{Gas modelling:} The employed \citet{springel03} hybrid multiphase interstellar medium model, and the implicit equilibrium assumption between the hot and cold phase might not be a good approximation for these central starbursts and lead to an underestimation of the stellar mass formed in them \citep{sparre2015}, thus not altering sizes enough compared to star-forming galaxies.
        \item \emph{AGN feedback modelling:} The model used in IllustrisTNG is by no means unique. While current cosmological volume simulations agree upon the need of AGN feedback to reproduce realistic $z=0$ galaxies, the way AGN feedback operates differs from simulation to simulation. The degree to which the presented results depend on AGN feedback modelling is an obvious next step to explore, and will either solidify the theoretical prediction, or open a new indirect probe into the nature of AGN feedback.
    \end{itemize}

In the coming years, progress in the available observational and modelling capabilities will enable further exploration of the relation between merger-induced starbursts and quenching in massive galaxies at $z\sim2$.

\acknowledgments
This research would not have been possible without the support of the Banneker Institute. We thank the anonymous referee for their helpful comments. We thank Wren Suess for the useful discussions and for sharing observational data from \citet{suess20} (included in Figure \ref{fig:halfmass}). We \textbf{also} thank Vicente Rodriguez-Gomez for sharing catalogues with more accessible merger information. S.B. acknowledges support from the Clay Fellowship. R.W. acknowledges support by Harvard University through the ITC Fellowship.

%

\vspace{5mm}


\software{astropy \citep{astropy}, numpy \citep{numpy}, scipy \citep{scipy}, matplotlib \citep{matplotlib}, h5py\footnote{\url{http://www.h5py.org}}.} \\




\bibliography{draft.bbl}{}

\begin{thebibliography}{}
\expandafter\ifx\csname natexlab\endcsname\relax\def\natexlab#1{#1}\fi
\providecommand{\url}[1]{\href{#1}{#1}}
\providecommand{\dodoi}[1]{doi:~\href{http://doi.org/#1}{\nolinkurl{#1}}}
\providecommand{\doeprint}[1]{\href{http://ascl.net/#1}{\nolinkurl{http://ascl.net/#1}}}
\providecommand{\doarXiv}[1]{\href{https://arxiv.org/abs/#1}{\nolinkurl{https://arxiv.org/abs/#1}}}

\bibitem[{{Abramson} {et~al.}(2014){Abramson}, {Kelson}, {Dressler},
  {Poggianti}, {Gladders}, {Oemler}, \& {Vulcani}}]{abramson14}
{Abramson}, L.~E., {Kelson}, D.~D., {Dressler}, A., {et~al.} 2014, \apjl, 785,
  L36, \dodoi{10.1088/2041-8205/785/2/L36}

\bibitem[{{Akhshik} {et~al.}(2020){Akhshik}, {Whitaker}, {Brammer}, {Mahler},
  {Sharon}, {Leja}, {Bayliss}, {Bezanson}, {Gladders}, {Man}, {Nelson},
  {Rigby}, {Rizzo}, {Toft}, {Wellons}, \& {Williams}}]{akhshik20}
{Akhshik}, M., {Whitaker}, K.~E., {Brammer}, G., {et~al.} 2020, \apj, 900, 184,
  \dodoi{10.3847/1538-4357/abac62}

\bibitem[{{Almaini} {et~al.}(2017){Almaini}, {Wild}, {Maltby}, {Hartley},
  {Simpson}, {Hatch}, {McLure}, {Dunlop}, \& {Rowlands}}]{2017MNRAS.472.1401A}
{Almaini}, O., {Wild}, V., {Maltby}, D.~T., {et~al.} 2017, \mnras, 472, 1401,
  \dodoi{10.1093/mnras/stx1957}

\bibitem[{{Astropy Collaboration} {et~al.}(2013){Astropy Collaboration},
  {Robitaille}, {Tollerud}, {Greenfield}, {Droettboom}, {Bray}, {Aldcroft},
  {Davis}, {Ginsburg}, {Price-Whelan}, {Kerzendorf}, {Conley}, {Crighton},
  {Barbary}, {Muna}, {Ferguson}, {Grollier}, {Parikh}, {Nair}, {Unther},
  {Deil}, {Woillez}, {Conseil}, {Kramer}, {Turner}, {Singer}, {Fox}, {Weaver},
  {Zabalza}, {Edwards}, {Azalee Bostroem}, {Burke}, {Casey}, {Crawford},
  {Dencheva}, {Ely}, {Jenness}, {Labrie}, {Lim}, {Pierfederici}, {Pontzen},
  {Ptak}, {Refsdal}, {Servillat}, \& {Streicher}}]{astropy}
{Astropy Collaboration}, {Robitaille}, T.~P., {Tollerud}, E.~J., {et~al.} 2013,
  \aap, 558, A33, \dodoi{10.1051/0004-6361/201322068}

\bibitem[{{Belli} {et~al.}(2015){Belli}, {Newman}, \&
  {Ellis}}]{2015ApJ...799..206B}
{Belli}, S., {Newman}, A.~B., \& {Ellis}, R.~S. 2015, \apj, 799, 206,
  \dodoi{10.1088/0004-637X/799/2/206}

\bibitem[{{Belli} {et~al.}(2019){Belli}, {Newman}, \&
  {Ellis}}]{2019ApJ...874...17B}
---. 2019, \apj, 874, 17, \dodoi{10.3847/1538-4357/ab07af}

\bibitem[{{Bower} {et~al.}(2006){Bower}, {Benson}, {Malbon}, {Helly}, {Frenk},
  {Baugh}, {Cole}, \& {Lacey}}]{2006MNRAS.370..645B}
{Bower}, R.~G., {Benson}, A.~J., {Malbon}, R., {et~al.} 2006, \mnras, 370, 645,
  \dodoi{10.1111/j.1365-2966.2006.10519.x}

\bibitem[{{Bruzual} \& {Charlot}(2003)}]{bruzual03}
{Bruzual}, G., \& {Charlot}, S. 2003, \mnras, 344, 1000,
  \dodoi{10.1046/j.1365-8711.2003.06897.x}

\bibitem[{{Choi} {et~al.}(2018){Choi}, {Somerville}, {Ostriker}, {Naab}, \&
  {Hirschmann}}]{choi18}
{Choi}, E., {Somerville}, R.~S., {Ostriker}, J.~P., {Naab}, T., \&
  {Hirschmann}, M. 2018, \apj, 866, 91, \dodoi{10.3847/1538-4357/aae076}

\bibitem[{{Croton} {et~al.}(2006){Croton}, {Springel}, {White}, {De Lucia},
  {Frenk}, {Gao}, {Jenkins}, {Kauffmann}, {Navarro}, \&
  {Yoshida}}]{2006MNRAS.365...11C}
{Croton}, D.~J., {Springel}, V., {White}, S. D.~M., {et~al.} 2006, \mnras, 365,
  11, \dodoi{10.1111/j.1365-2966.2005.09675.x}

\bibitem[{{Dekel} \& {Burkert}(2014)}]{2014MNRAS.438.1870D}
{Dekel}, A., \& {Burkert}, A. 2014, \mnras, 438, 1870,
  \dodoi{10.1093/mnras/stt2331}

\bibitem[{{D'Eugenio} {et~al.}(2020){D'Eugenio}, {van der Wel}, {Wu}, {Barone},
  {van Houdt}, {Bezanson}, {Straatman}, {Pacifici}, {Muzzin}, {Gallazzi},
  {Wild}, {Sobral}, {Bell}, {Zibetti}, {Mowla}, \&
  {Franx}}]{2020MNRAS.497..389D}
{D'Eugenio}, F., {van der Wel}, A., {Wu}, P.-F., {et~al.} 2020, \mnras, 497,
  389, \dodoi{10.1093/mnras/staa1937}

\bibitem[{{Genel} {et~al.}(2018){Genel}, {Nelson}, {Pillepich}, {Springel},
  {Pakmor}, {Weinberger}, {Hernquist}, {Naiman}, {Vogelsberger}, {Marinacci},
  \& {Torrey}}]{genel18}
{Genel}, S., {Nelson}, D., {Pillepich}, A., {et~al.} 2018, \mnras, 474, 3976,
  \dodoi{10.1093/mnras/stx3078}

\bibitem[{Harris {et~al.}(2020)Harris, Millman, van~der Walt, Gommers,
  Virtanen, Cournapeau, Wieser, Taylor, Berg, Smith, Kern, Picus, Hoyer, van
  Kerkwijk, Brett, Haldane, del R{\'{i}}o, Wiebe, Peterson,
  G{\'{e}}rard-Marchant, Sheppard, Reddy, Weckesser, Abbasi, Gohlke, \&
  Oliphant}]{numpy}
Harris, C.~R., Millman, K.~J., van~der Walt, S.~J., {et~al.} 2020, Nature, 585,
  357, \dodoi{10.1038/s41586-020-2649-2}

\bibitem[{{Hunter}(2007)}]{matplotlib}
{Hunter}, J.~D. 2007, Computing in Science and Engineering, 9, 90,
  \dodoi{10.1109/MCSE.2007.55}

\bibitem[{{Jafariyazani} {et~al.}(2020){Jafariyazani}, {Newman}, {Mobasher},
  {Belli}, {Ellis}, \& {Patel}}]{jafariyazani20}
{Jafariyazani}, M., {Newman}, A.~B., {Mobasher}, B., {et~al.} 2020, \apjl, 897,
  L42, \dodoi{10.3847/2041-8213/aba11c}

\bibitem[{{Ludlow} {et~al.}(2019){Ludlow}, {Schaye}, {Schaller}, \&
  {Richings}}]{ludlow2019}
{Ludlow}, A.~D., {Schaye}, J., {Schaller}, M., \& {Richings}, J. 2019, \mnras,
  488, L123, \dodoi{10.1093/mnrasl/slz110}

\bibitem[{{Madau} \& {Dickinson}(2014)}]{2014ARA&A..52..415M}
{Madau}, P., \& {Dickinson}, M. 2014, \araa, 52, 415,
  \dodoi{10.1146/annurev-astro-081811-125615}

\bibitem[{{Marinacci} {et~al.}(2018){Marinacci}, {Vogelsberger}, {Pakmor},
  {Torrey}, {Springel}, {Hernquist}, {Nelson}, {Weinberger}, {Pillepich},
  {Naiman}, \& {Genel}}]{2018MNRAS.480.5113M}
{Marinacci}, F., {Vogelsberger}, M., {Pakmor}, R., {et~al.} 2018, \mnras, 480,
  5113, \dodoi{10.1093/mnras/sty2206}

\bibitem[{{Muzzin} {et~al.}(2013){Muzzin}, {Marchesini}, {Stefanon}, {Franx},
  {McCracken}, {Milvang-Jensen}, {Dunlop}, {Fynbo}, {Brammer}, {Labb{\'e}}, \&
  {van Dokkum}}]{muzzin13}
{Muzzin}, A., {Marchesini}, D., {Stefanon}, M., {et~al.} 2013, \apj, 777, 18,
  \dodoi{10.1088/0004-637X/777/1/18}

\bibitem[{{Naiman} {et~al.}(2018){Naiman}, {Pillepich}, {Springel},
  {Ramirez-Ruiz}, {Torrey}, {Vogelsberger}, {Pakmor}, {Nelson}, {Marinacci},
  {Hernquist}, {Weinberger}, \& {Genel}}]{2018MNRAS.477.1206N}
{Naiman}, J.~P., {Pillepich}, A., {Springel}, V., {et~al.} 2018, \mnras, 477,
  1206, \dodoi{10.1093/mnras/sty618}

\bibitem[{{Nelson} {et~al.}(2018){Nelson}, {Pillepich}, {Springel},
  {Weinberger}, {Hernquist}, {Pakmor}, {Genel}, {Torrey}, {Vogelsberger},
  {Kauffmann}, {Marinacci}, \& {Naiman}}]{2018MNRAS.475..624N}
{Nelson}, D., {Pillepich}, A., {Springel}, V., {et~al.} 2018, \mnras, 475, 624,
  \dodoi{10.1093/mnras/stx3040}

\bibitem[{{Nelson} {et~al.}(2019){Nelson}, {Springel}, {Pillepich},
  {Rodriguez-Gomez}, {Torrey}, {Genel}, {Vogelsberger}, {Pakmor}, {Marinacci},
  {Weinberger}, {Kelley}, {Lovell}, {Diemer}, \& {Hernquist}}]{nelson19}
{Nelson}, D., {Springel}, V., {Pillepich}, A., {et~al.} 2019, Computational
  Astrophysics and Cosmology, 6, 2, \dodoi{10.1186/s40668-019-0028-x}

\bibitem[{{Nelson} {et~al.}(2016){Nelson}, {van Dokkum}, {F{\"o}rster
  Schreiber}, {Franx}, {Brammer}, {Momcheva}, {Wuyts}, {Whitaker}, {Skelton},
  {Fumagalli}, {Hayward}, {Kriek}, {Labb{\'e}}, {Leja}, {Rix}, {Tacconi}, {van
  der Wel}, {van den Bosch}, {Oesch}, {Dickey}, \& {Ulf Lange}}]{nelson16}
{Nelson}, E.~J., {van Dokkum}, P.~G., {F{\"o}rster Schreiber}, N.~M., {et~al.}
  2016, \apj, 828, 27, \dodoi{10.3847/0004-637X/828/1/27}

\bibitem[{{Nelson} {et~al.}(2021){Nelson}, {Tacchella}, {Diemer}, {Leja},
  {Hernquist}, {Whitaker}, {Weinberger}, {Pillepich}, {Nelson}, {Terrazas},
  {Nevin}, {Brammer}, {Burkhart}, {Cochrane}, {van Dokkum}, {Johnson}, {Mowla},
  {Pakmor}, {Skelton}, {Speagle}, {Springel}, {Torrey}, {Vogelsberger}, \&
  {Wuyts}}]{nelson21}
{Nelson}, E.~J., {Tacchella}, S., {Diemer}, B., {et~al.} 2021, arXiv e-prints,
  arXiv:2101.12212.
\newblock \doarXiv{2101.12212}

\bibitem[{Pathak {et~al.}(2021)Pathak, Weinberger, Belli, Suess, \&
  Rodriguez-Gomez}]{zenodo}
Pathak, D., Weinberger, R., Belli, S., Suess, K.~A., \& Rodriguez-Gomez, V.
  2021, pathakde/Pathak\_2021: Pathak et al. 2021, v1.0,  Zenodo,
  \dodoi{10.5281/zenodo.5093727}

\bibitem[{{Pillepich} {et~al.}(2018{\natexlab{a}}){Pillepich}, {Nelson},
  {Hernquist}, {Springel}, {Pakmor}, {Torrey}, {Weinberger}, {Genel}, {Naiman},
  {Marinacci}, \& {Vogelsberger}}]{2018MNRAS.475..648P}
{Pillepich}, A., {Nelson}, D., {Hernquist}, L., {et~al.} 2018{\natexlab{a}},
  \mnras, 475, 648, \dodoi{10.1093/mnras/stx3112}

\bibitem[{{Pillepich} {et~al.}(2018{\natexlab{b}}){Pillepich}, {Springel},
  {Nelson}, {Genel}, {Naiman}, {Pakmor}, {Hernquist}, {Torrey}, {Vogelsberger},
  {Weinberger}, \& {Marinacci}}]{2018MNRAS.473.4077P}
{Pillepich}, A., {Springel}, V., {Nelson}, D., {et~al.} 2018{\natexlab{b}},
  \mnras, 473, 4077, \dodoi{10.1093/mnras/stx2656}

\bibitem[{{Planck Collaboration} {et~al.}(2016){Planck Collaboration}, Ade, \&
  Aghanim}]{Ade2016}
{Planck Collaboration}, Ade, P. A.~R., \& Aghanim, N. e.~a. 2016, Astronomy \&
  Astrophysics, 594, \dodoi{10.1051/0004-6361/201525830}

\bibitem[{{Pontzen} {et~al.}(2017){Pontzen}, {Tremmel}, {Roth}, {Peiris},
  {Saintonge}, {Volonteri}, {Quinn}, \& {Governato}}]{pontzen17}
{Pontzen}, A., {Tremmel}, M., {Roth}, N., {et~al.} 2017, \mnras, 465, 547,
  \dodoi{10.1093/mnras/stw2627}

\bibitem[{{Rodriguez-Gomez} {et~al.}(2015){Rodriguez-Gomez}, {Genel},
  {Vogelsberger}, {Sijacki}, {Pillepich}, {Sales}, {Torrey}, {Snyder},
  {Nelson}, {Springel}, {Ma}, \& {Hernquist}}]{2015MNRAS.449...49R}
{Rodriguez-Gomez}, V., {Genel}, S., {Vogelsberger}, M., {et~al.} 2015, \mnras,
  449, 49, \dodoi{10.1093/mnras/stv264}

\bibitem[{{Sanchez} {et~al.}(2021){Sanchez}, {Tremmel}, {Werk}, {Pontzen},
  {Christensen}, {Quinn}, {Loebman}, \& {Cruz}}]{sanchez21}
{Sanchez}, N.~N., {Tremmel}, M., {Werk}, J.~K., {et~al.} 2021, \apj, 911, 116,
  \dodoi{10.3847/1538-4357/abeb15}

\bibitem[{Schawinski {et~al.}(2014)Schawinski, Urry, Simmons, Fortson, Kaviraj,
  Keel, Lintott, Masters, Nichol, Sarzi, Skibba, Treister, Willett, Wong, \&
  Yi}]{10.1093/mnras/stu327}
Schawinski, K., Urry, C.~M., Simmons, B.~D., {et~al.} 2014, Monthly Notices of
  the Royal Astronomical Society, 440, 889, \dodoi{10.1093/mnras/stu327}

\bibitem[{{Setton} {et~al.}(2020){Setton}, {Bezanson}, {Suess}, {Hunt},
  {Greene}, {Kriek}, {Spilker}, {Feldmann}, \& {Narayanan}}]{setton20}
{Setton}, D.~J., {Bezanson}, R., {Suess}, K.~A., {et~al.} 2020, \apj, 905, 79,
  \dodoi{10.3847/1538-4357/abc265}

\bibitem[{{Sharma} {et~al.}(2021){Sharma}, {Choi}, {Somerville}, {Snyder},
  {Kocevski}, {Hirschmann}, {Moster}, {Naab}, {Narayanan}, {Ostriker}, \&
  {Rosario}}]{sharma21}
{Sharma}, R.~S., {Choi}, E., {Somerville}, R.~S., {et~al.} 2021, arXiv
  e-prints, arXiv:2101.01729.
\newblock \doarXiv{2101.01729}

\bibitem[{{Sparre} {et~al.}(2015){Sparre}, {Hayward}, {Springel},
  {Vogelsberger}, {Genel}, {Torrey}, {Nelson}, {Sijacki}, \&
  {Hernquist}}]{sparre2015}
{Sparre}, M., {Hayward}, C.~C., {Springel}, V., {et~al.} 2015, \mnras, 447,
  3548, \dodoi{10.1093/mnras/stu2713}

\bibitem[{{Speagle} {et~al.}(2014){Speagle}, {Steinhardt}, {Capak}, \&
  {Silverman}}]{speagle14}
{Speagle}, J.~S., {Steinhardt}, C.~L., {Capak}, P.~L., \& {Silverman}, J.~D.
  2014, \apjs, 214, 15, \dodoi{10.1088/0067-0049/214/2/15}

\bibitem[{{Springel} \& {Hernquist}(2003)}]{springel03}
{Springel}, V., \& {Hernquist}, L. 2003, \mnras, 339, 289,
  \dodoi{10.1046/j.1365-8711.2003.06206.x}

\bibitem[{{Springel} {et~al.}(2001){Springel}, {White}, {Tormen}, \&
  {Kauffmann}}]{2001MNRAS.328..726S}
{Springel}, V., {White}, S. D.~M., {Tormen}, G., \& {Kauffmann}, G. 2001,
  \mnras, 328, 726, \dodoi{10.1046/j.1365-8711.2001.04912.x}

\bibitem[{{Springel} {et~al.}(2018){Springel}, {Pakmor}, {Pillepich},
  {Weinberger}, {Nelson}, {Hernquist}, {Vogelsberger}, {Genel}, {Torrey},
  {Marinacci}, \& {Naiman}}]{2018MNRAS.475..676S}
{Springel}, V., {Pakmor}, R., {Pillepich}, A., {et~al.} 2018, \mnras, 475, 676,
  \dodoi{10.1093/mnras/stx3304}

\bibitem[{{Suess} {et~al.}(2020){Suess}, {Kriek}, {Price}, \&
  {Barro}}]{suess20}
{Suess}, K.~A., {Kriek}, M., {Price}, S.~H., \& {Barro}, G. 2020, \apjl, 899,
  L26, \dodoi{10.3847/2041-8213/abacc9}

\bibitem[{{Suess} {et~al.}(2021){Suess}, {Kriek}, {Price}, \&
  {Barro}}]{suess21}
---. 2021, arXiv e-prints, arXiv:2101.05820.
\newblock \doarXiv{2101.05820}

\bibitem[{{Tacchella} {et~al.}(2016){Tacchella}, {Dekel}, {Carollo},
  {Ceverino}, {DeGraf}, {Lapiner}, {Mandelker}, \& {Primack
  Joel}}]{tacchella16}
{Tacchella}, S., {Dekel}, A., {Carollo}, C.~M., {et~al.} 2016, \mnras, 457,
  2790, \dodoi{10.1093/mnras/stw131}

\bibitem[{{Tacchella} {et~al.}(2015){Tacchella}, {Carollo}, {Renzini},
  {F{\"o}rster Schreiber}, {Lang}, {Wuyts}, {Cresci}, {Dekel}, {Genzel},
  {Lilly}, {Mancini}, {Newman}, {Onodera}, {Shapley}, {Tacconi}, {Woo}, \&
  {Zamorani}}]{tacchella15}
{Tacchella}, S., {Carollo}, C.~M., {Renzini}, A., {et~al.} 2015, Science, 348,
  314, \dodoi{10.1126/science.1261094}

\bibitem[{{Tacchella} {et~al.}(2019){Tacchella}, {Diemer}, {Hernquist},
  {Genel}, {Marinacci}, {Nelson}, {Pillepich}, {Rodriguez-Gomez}, {Sales},
  {Springel}, \& {Vogelsberger}}]{tacchella19}
{Tacchella}, S., {Diemer}, B., {Hernquist}, L., {et~al.} 2019, \mnras, 487,
  5416, \dodoi{10.1093/mnras/stz1657}

\bibitem[{{Tadaki} {et~al.}(2020){Tadaki}, {Belli}, {Burkert}, {Dekel},
  {F{\"o}rster Schreiber}, {Genzel}, {Hayashi}, {Herrera-Camus}, {Kodama},
  {Kohno}, {Koyama}, {Lee}, {Lutz}, {Mowla}, {Nelson}, {Renzini}, {Suzuki},
  {Tacconi}, {{\"U}bler}, {Wisnioski}, \& {Wuyts}}]{tadaki20}
{Tadaki}, K.-i., {Belli}, S., {Burkert}, A., {et~al.} 2020, \apj, 901, 74,
  \dodoi{10.3847/1538-4357/abaf4a}

\bibitem[{Virtanen {et~al.}(2020)Virtanen, Gommers, Oliphant, Haberland, Reddy,
  Cournapeau, Burovski, Peterson, Weckesser, Bright, {van der Walt}, Brett,
  Wilson, Millman, Mayorov, Nelson, Jones, Kern, Larson, Carey, Polat, Feng,
  Moore, {VanderPlas}, Laxalde, Perktold, Cimrman, Henriksen, Quintero, Harris,
  Archibald, Ribeiro, Pedregosa, {van Mulbregt}, \& {SciPy 1.0
  Contributors}}]{scipy}
Virtanen, P., Gommers, R., Oliphant, T.~E., {et~al.} 2020, Nature Methods, 17,
  261, \dodoi{10.1038/s41592-019-0686-2}

\bibitem[{{Vogelsberger} {et~al.}(2013){Vogelsberger}, {Genel}, {Sijacki},
  {Torrey}, {Springel}, \& {Hernquist}}]{vogelsberger13}
{Vogelsberger}, M., {Genel}, S., {Sijacki}, D., {et~al.} 2013, \mnras, 436,
  3031, \dodoi{10.1093/mnras/stt1789}

\bibitem[{{Weinberger} {et~al.}(2017){Weinberger}, {Springel}, {Hernquist},
  {Pillepich}, {Marinacci}, {Pakmor}, {Nelson}, {Genel}, {Vogelsberger},
  {Naiman}, \& {Torrey}}]{2017MNRAS.465.3291W}
{Weinberger}, R., {Springel}, V., {Hernquist}, L., {et~al.} 2017, \mnras, 465,
  3291, \dodoi{10.1093/mnras/stw2944}

\bibitem[{{Weinberger} {et~al.}(2018){Weinberger}, {Springel}, {Pakmor},
  {Nelson}, {Genel}, {Pillepich}, {Vogelsberger}, {Marinacci}, {Naiman},
  {Torrey}, \& {Hernquist}}]{weinberger18}
{Weinberger}, R., {Springel}, V., {Pakmor}, R., {et~al.} 2018, \mnras, 479,
  4056, \dodoi{10.1093/mnras/sty1733}

\bibitem[{{Whitaker} {et~al.}(2012){Whitaker}, {Kriek}, {van Dokkum},
  {Bezanson}, {Brammer}, {Franx}, \& {Labb{\'e}}}]{2012ApJ...745..179W}
{Whitaker}, K.~E., {Kriek}, M., {van Dokkum}, P.~G., {et~al.} 2012, \apj, 745,
  179, \dodoi{10.1088/0004-637X/745/2/179}

\bibitem[{{Wild} {et~al.}(2016){Wild}, {Almaini}, {Dunlop}, {Simpson},
  {Rowlands}, {Bowler}, {Maltby}, \& {McLure}}]{2016MNRAS.463..832W}
{Wild}, V., {Almaini}, O., {Dunlop}, J., {et~al.} 2016, \mnras, 463, 832,
  \dodoi{10.1093/mnras/stw1996}

\bibitem[{{Wild} {et~al.}(2009){Wild}, {Walcher}, {Johansson}, {Tresse},
  {Charlot}, {Pollo}, {Le F{\`e}vre}, \& {de Ravel}}]{wild09}
{Wild}, V., {Walcher}, C.~J., {Johansson}, P.~H., {et~al.} 2009, \mnras, 395,
  144, \dodoi{10.1111/j.1365-2966.2009.14537.x}

\bibitem[{{Wu} {et~al.}(2018){Wu}, {van der Wel}, {Bezanson}, {Gallazzi},
  {Pacifici}, {Straatman}, {Bari{\v{s}}i{\'c}}, {Bell}, {Chauke}, {van Houdt},
  {Franx}, {Muzzin}, {Sobral}, \& {Wild}}]{2018ApJ...868...37W}
{Wu}, P.-F., {van der Wel}, A., {Bezanson}, R., {et~al.} 2018, \apj, 868, 37,
  \dodoi{10.3847/1538-4357/aae822}

\bibitem[{{Wu} {et~al.}(2020){Wu}, {van der Wel}, {Bezanson}, {Gallazzi},
  {Pacifici}, {Straatman}, {Bari{\v{s}}i{\'c}}, {Bell}, {Chauke}, {D'Eugenio},
  {Franx}, {Muzzin}, {Sobral}, \& {van Houdt}}]{wu20}
---. 2020, \apj, 888, 77, \dodoi{10.3847/1538-4357/ab5fd9}

\bibitem[{{Yano} {et~al.}(2016){Yano}, {Kriek}, {van der Wel}, \&
  {Whitaker}}]{2016ApJ...817L..21Y}
{Yano}, M., {Kriek}, M., {van der Wel}, A., \& {Whitaker}, K.~E. 2016, \apjl,
  817, L21, \dodoi{10.3847/2041-8205/817/2/L21}

\bibitem[{{Zolotov} {et~al.}(2015){Zolotov}, {Dekel}, {Mandelker}, {Tweed},
  {Inoue}, {DeGraf}, {Ceverino}, {Primack}, {Barro}, \& {Faber}}]{zolotov15}
{Zolotov}, A., {Dekel}, A., {Mandelker}, N., {et~al.} 2015, \mnras, 450, 2327,
  \dodoi{10.1093/mnras/stv740}

\end{thebibliography}
\bibliographystyle{aasjournal}



\end{document}